\documentclass[prl,twocolumn,showpacs,superscriptaddress,floatfix]{revtex4}

\usepackage{graphicx}

 \usepackage{xspace}

\begin{document}

\newcommand{\E}{\mathrm{e}}
\newcommand{\I}{\rm{i}}
\newcommand{\su}[1]{\mathcal{#1}}
\newcommand{\dt}{\: \frac{d}{d t}}
\newcommand{\rf}[1]{Eq.~(\ref{#1})}
\newcommand{\rfg}[1]{Fig.~\ref{#1}}
\newcommand{\ex}[1]{\langle #1 \rangle}
\newcommand{\da}{^\dagger}
\newcommand{\pda}{^{\phantom{\dagger}}}
\newcommand{\wlo}{\omega_{\rm{LO}}}
\newcommand{\ps}{^{\phantom{*}}}
\newcommand{\bk}{\mathbf{k}}
\newcommand{\bq}{\mathbf{q}}
\newcommand{\bQ}{\mathbf{Q}}
\newcommand{\br}{\mathbf{r}}
\newcommand{\pa}{_{||}}
\newcommand{\pe}{_\perp}
 \newcommand{\bE}{\mathbf{\rm E}}
\newcommand{\bP}{\mathbf{P}}
\newcommand{\ubq }{\underline{\mathbf{q}}}

 \newcommand{\bks}{\mathbf{k'}}
\def\inprint{in print}
\title{Exceeding the Manley-Rowe quantum efficiency limit in an optically pumped THz amplifier}

\author{In\`es Waldmueller}\email{iwaldmu@sandia.gov}
\affiliation{Sandia National Laboratories
Albuquerque, New Mexico 87185-1086}
\author{Michael C. Wanke}
\affiliation{Sandia National Laboratories
Albuquerque, New Mexico 87185-1086}
\author{ Weng W. Chow}
\affiliation{Sandia National Laboratories
Albuquerque, New Mexico 87185-1086}
\affiliation{Physics Department and Institute of Quantum Studies, Texas A \& M University, College Station, Texas 77843}

\date{\today}

\begin{abstract}
\noindent Using a microscopic theory based on the Maxwell-semiconductor Bloch equations, we investigate the possibility of an optically-assisted electrically-driven THz quantum cascade laser. 
Whereas in optical conversion schemes the power conversion efficiency is limited by the Manley-Rowe relation, the proposed optically-assisted scheme can achieve higher efficiency by coherently recovering the optical pump energy. Furthermore, due to quantum coherence effects the detrimental effects of scattering are mitigated.
\end{abstract}

\maketitle

Quantum cascade lasers (QCLs) have become an important topic during the last decade. Based on transitions between conduction subbands, QCLs have been fabricated for a wide range of infrared frequencies. Recently, significant interest has focused on the development of QCLs in the THz regime \cite{kohler:2002}. 
In a THz-QCL, carriers are injected directly into the energetically higher laser subband (direct THz-QCL). The lower laser subband can be depleted by zero, single or double optical-phonon scattering, \cite{kohler:2002},\cite{williams:2003},\cite{williams:2006}, yielding a population inversion between the lasing subbands. However, the required threshold is reasonably low only at low temperature. As both lasing subbands are located energetically low in the structure [see Fig. 1 (a)], thermal backfilling destroys the population inversion at high temperature. Additionally, temperature dependent scattering processes (parasitic current channels) also reduce the population inversion. As a result, lasing threshold of direct QCLs in the THZ regime increases appreciably with increasing temperature, limiting the present maximum operation temperature to 164 K \cite{kumar:2004}. 

Optical conversion is one approach to circumvent the problem of thermal backfilling, e.g.\cite{gauthier:1999},\cite{liu_raman:2003}. Instead of electrically injecting the carriers directly into the upper laser subband, the upper laser subband is populated by an external optical field [see Fig. 1(b)]. However, while conventional optical conversion presents a solution to the problem of population thermalization, it also introduces a fundamental constraint due to the Manley-Rowe quantum limit, i.e. the highest achievable conversion efficiency is given by the quotient of output and input frequencies. 
\begin{figure}
\vspace*{-1cm}$\:$\\\hspace*{-1cm}\includegraphics[width=0.9\columnwidth,angle=270]{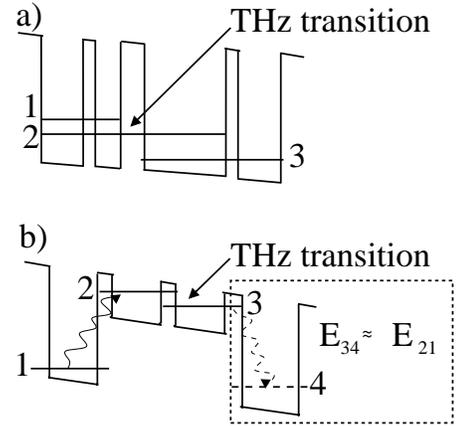}\vspace*{-1cm}$\:$\\
 \caption{a) Single stage of a direct THz-QCL: carriers are injected in subband 1 and extracted from subband 3. b) Single stage of proposed optically-assisted electrically-driven scheme. Without the pump recovery portion (inside dashed box) the scheme reverts to a conventional optical conversion configuration.\vspace*{-1cm}$\:$\\}
 \end{figure}

In this letter, we propose and model an optically-assisted electrically-driven quantum cascade laser. In contrast to optical conversion schemes, the THz energy is not derived from the external optical field, but comes from the forward electrical bias as in a conventional QCL, thus allowing coherently recovery the pump energy.  In this way we get the advantages of an optical conversion scheme,  but for first time are not constrained by the Manley-Rowe limit. The basic idea is sketched in Fig. 1 (b). Crucial to the approach is the design of the coupled quantum wells so that the transition frequencies between subbands 1 and 2, and subbands 3 and 4 are sufficiently similar so that an external optical field can simultaneously populate the higher laser subband (2) and deplete the lower laser subband (3), thus regaining the pump photons via stimulated emission. 

We investigate our scheme theoretically using a microscopic approach that considers the nonlinear interactions between the externally applied field, the THz-field, and the active region. The active medium is treated within the framework of the semiconductor Bloch equations for polarizations $P_{ij}(x,t)$ between subbands $i$ and $j$ and subband populations $n_i(x,t)$ at time $t$ and in-plane direction $x$ (in RWA):
\begin{widetext}
\begin{eqnarray}
\dt P_{12}(x,t) &=& \frac{\I}{\hbar} ({\epsilon}_{23}  +\hbar \omega_{\text{D}}) P_{12}  - \I (n_2(x,t)-n_1(x,t)) \Omega_{\text{D}}(x,t) + \I P_{13} \Omega_{\text{P}}(x,t) - \gamma  P_{12}\\
\dt P_{23}(x,t) &=& \frac{\I}{\hbar} ({\epsilon}_{23}  - \hbar \omega_{\text{P}} ) P_{23}  +\I (n_2(x,t)-n_3(x,t)) \Omega^*_{\text{P}}(x,t) - \I P_{13} \Omega^*_{\text{D}}(x,t) + \I P_{24} \Omega_{\text{D}}(x,t) - \gamma  P_{23} \\
\dt P_{34}(x,t) &=& \frac{\I}{\hbar} ({\epsilon}_{34}  -\hbar \omega_{\text{D}} ) P_{34}  + \I (n_3(x,t)-n_4(x,t)) \Omega^*_{\text{D}}(x,t) - \I P_{24} \Omega_{\text{P}}(x,t) - \gamma  P_{34} \\
\dt P_{13}(x,t) &=& \frac{\I}{\hbar} ({\epsilon}_{13}  +\hbar \omega_{\text{D}} - \hbar \omega_{\text{P}} ) P_{13}  + \I P_{12} \Omega^*_{\text{P}}(x,t) - \I (P_{23}-P_{14}) \Omega_{\text{D}}(x,t) - \gamma  P_{13} \\
\dt P_{14}(x,t) &=& \frac{\I}{\hbar} ({\epsilon}_{14}  - \hbar \omega_{\text{P}} ) P_{14}  - \I P_{24} \Omega_{\text{D}}(x,t) + \I P_{13} \Omega^*_{\text{D}}(x,t) - \gamma  P_{14} \\
\dt P_{24}(x,t) &=& \frac{\I}{\hbar} ({\epsilon}_{24}  -\hbar \omega_{\text{D}} - \hbar \omega_{\text{P}} ) P_{24}  + \I (P_{23}-P_{14}) \Omega_{\text{D}}(x,t) - \I P_{34} \Omega^*_{\text{P}}(x,t) - \gamma  P_{24} \\
\dt n_1(x,t) &=& 2 \text{Im}(P^*_{12}(x,t) \Omega_{\text{D}}(x,t)) + \dt n_1|_{\text{scatt}} \\
\dt n_2(x,t) &=& 2 \text{Im}(P_{12}(x,t) \Omega^*_{\text{D}}(x,t)) + 2 \text{Im}(P^*_{23}(x,t) \Omega^*_{\text{P}}(x,t)) + \dt n_2|_{\text{scatt}} \\
\dt n_3(x,t) &=& 2 \text{Im}(P_{23}(x,t) \Omega^*_{\text{P}}(x,t)) + 2 \text{Im}(P^*_{34}(x,t) \Omega^*_{\text{D}}(x,t))+ \dt n_3|_{\text{scatt}} \\
\dt n_4(x,t) &=& 2 \text{Im}(P_{34}(x,t) \Omega_{\text{D}}(x,t)) + \dt n_4|_{\text{scatt}} 
\end{eqnarray}
\end{widetext}
where ${\epsilon}_{ij}$ is the transition energy between subbands $i$ and $j$. $\Omega_{\text{D,P}}(t) = 1/(2 \hbar) (-e d_{12,23} \su{E_{\text{D,P}}}(t))$ is the Rabi frequency for the driving field (D) or the THz-field (P) with dipole moment $d_{ij}$. 
For our configuration we calculated $d_{12} \approx d_{34} \approx 0.2 \: d_{23}$. In Eqs. 1-6, the terms containing the subband populations account for stimulated emission or absorption, while the terms that are products of the Rabi frequencies and polarizations describe quantum coherence effects. In contrast to dipole-allowed ($d\neq 0$) polarizations, the dipole-forbidden ($d\equiv 0$) or dipole-suppressed ($d\approx 0$) polarizations, $P_{13},P_{14}$ and $P_{24}$, are only driven by quantum coherence contributions. $\gamma$ denotes the dephasing rate of the polarizations and $\dt n_i|_{\text{scatt}}$ accounts for population relaxation effects due to carrier-carrier and carrier-phonon scattering. These contributions are taken into account in the relaxation rate approximation with microscopically determined scattering rates [for details see e.g. \cite{we:IEEE}]. 
$\su{E_{\text{a}}}(x,t)$ (a = D,P) denotes the complex field amplitude of the externally applied field (a=D) and the THz-field (a=P) propagating in the x-direction which are determined by solving the reduced wave equations in the retarded time frame:
\begin{eqnarray}
\hspace*{-0.5cm} &&E_a(x,t) = \frac{1}{2} \: \su{E_{\text{a}}}(x,t) \: u_a(y,z) \:\E^{-\I \omega_a(t-x/c \: n(y,z) )} +c.c.  \:, \qquad \\
\hspace*{-0.5cm}&&\frac{d}{dx} \su{E_{\text{a}}}(x,t) = {\I} \frac{\omega_a \Gamma_a}{c \epsilon_0 \sqrt{\epsilon_B} V} \sum_{i,j} d_{nm} P_{ij}(x,t) \E^{\I \omega_a t}  \:.
\end{eqnarray}
Here $\Gamma_a$ gives the 2D-confinement (y,z) of the field mode $a$ to the active region.
\begin{figure*}
\includegraphics[width=0.43\columnwidth,angle=270]{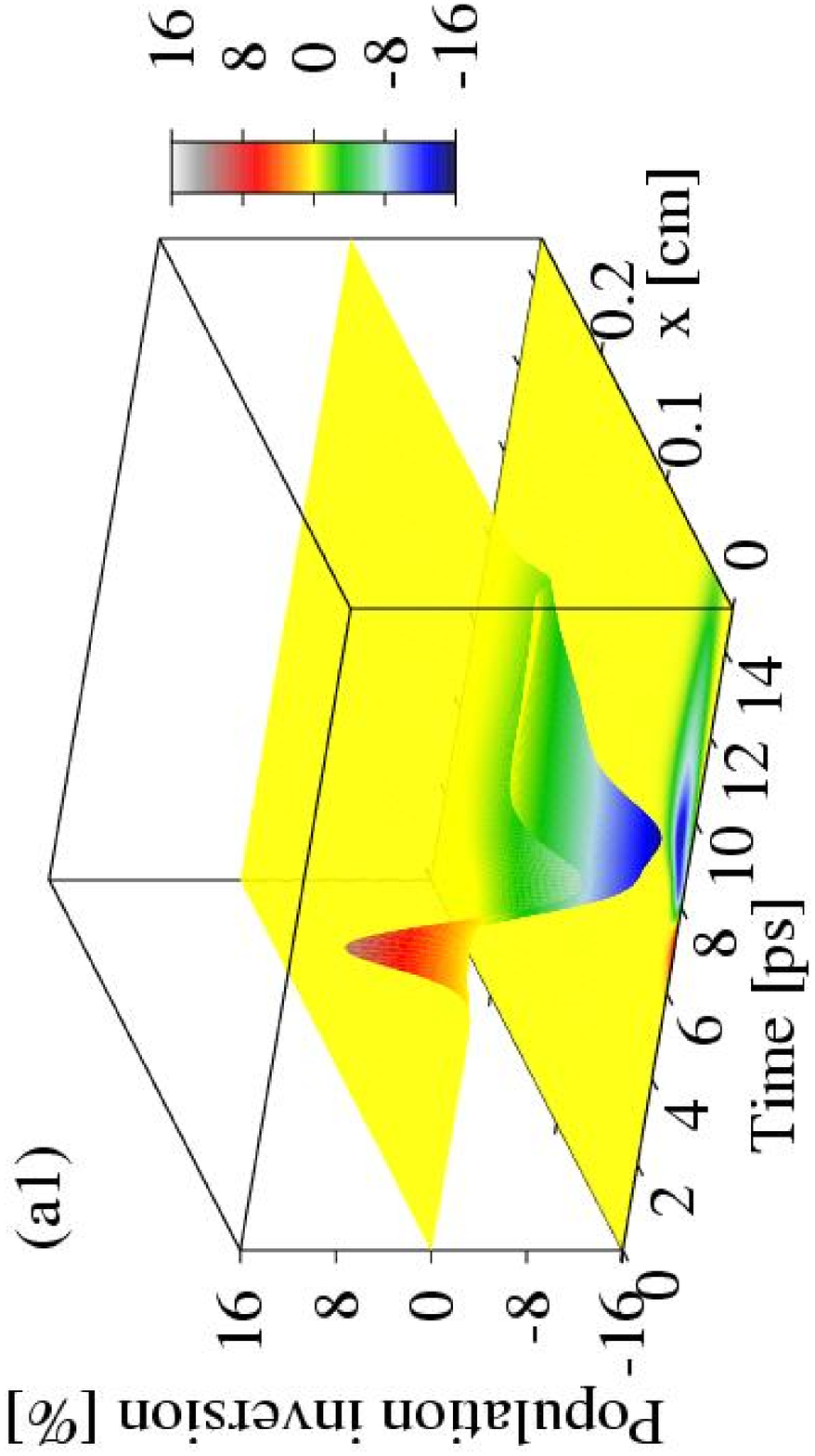}\hspace*{0.cm}
\includegraphics[width=0.43\columnwidth,angle=270]{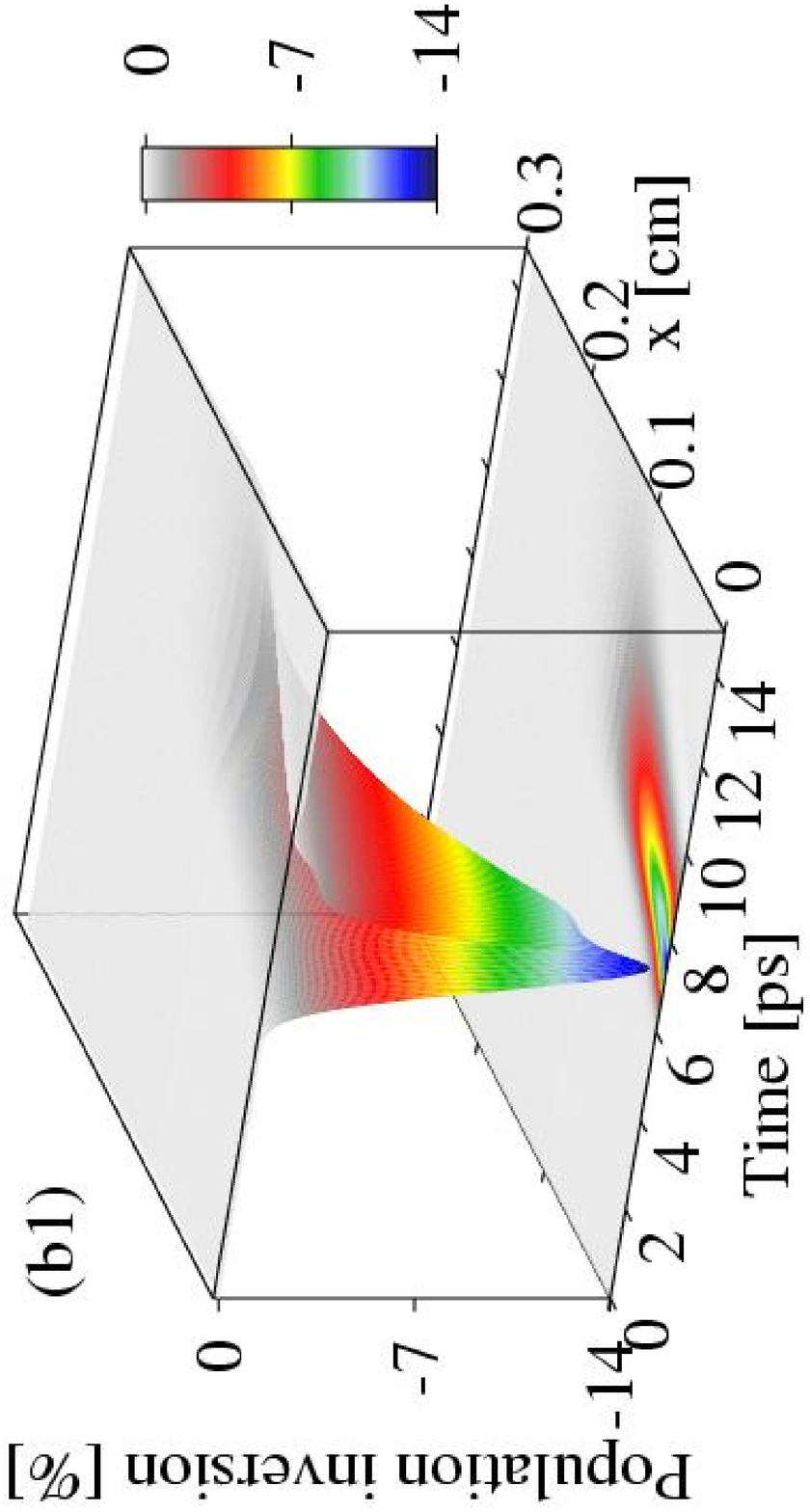}\hspace*{0.cm}
\includegraphics[width=0.43\columnwidth,angle=270]{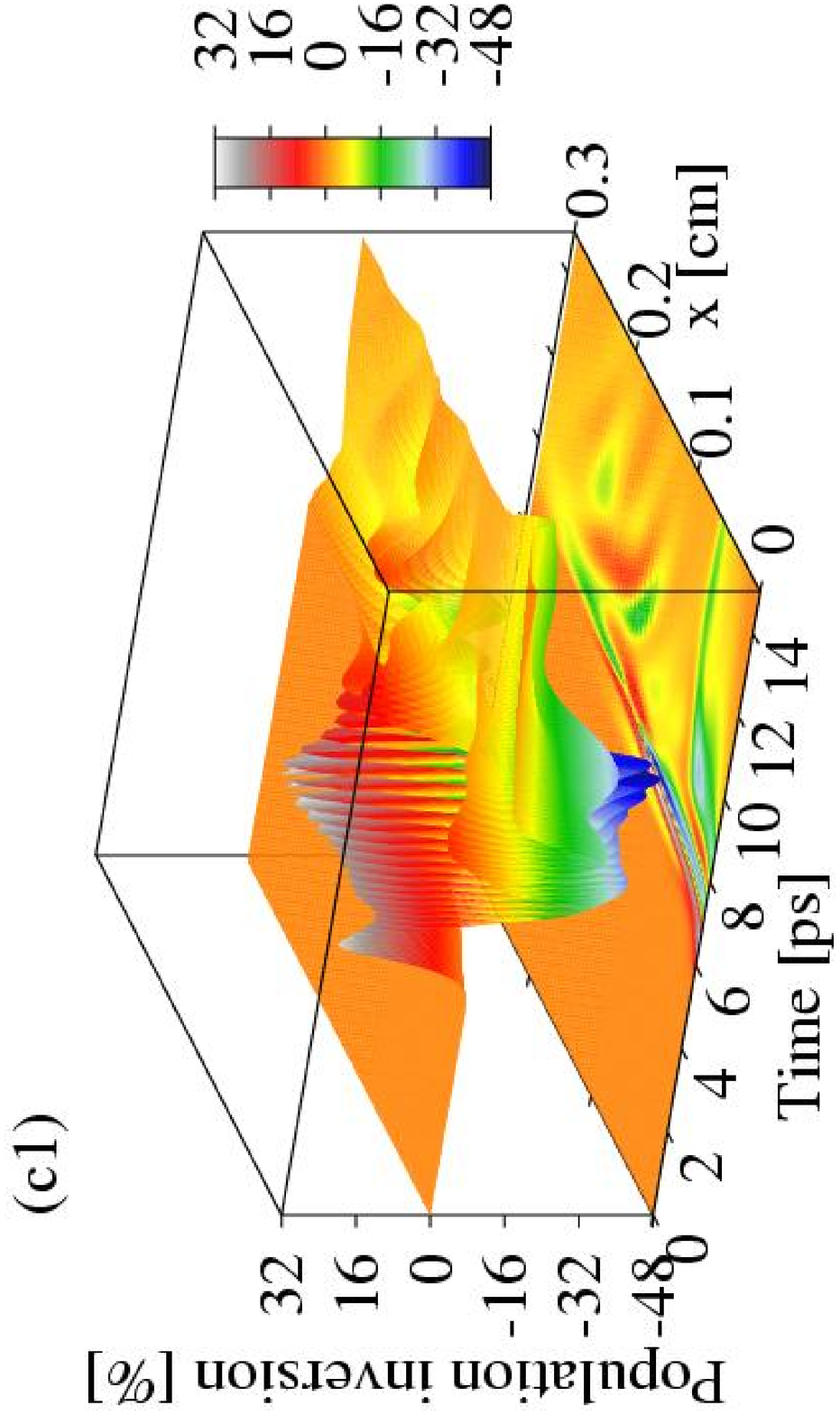} \vspace{-0.4cm}  \\ 
\includegraphics[width=0.43\columnwidth,angle=270]{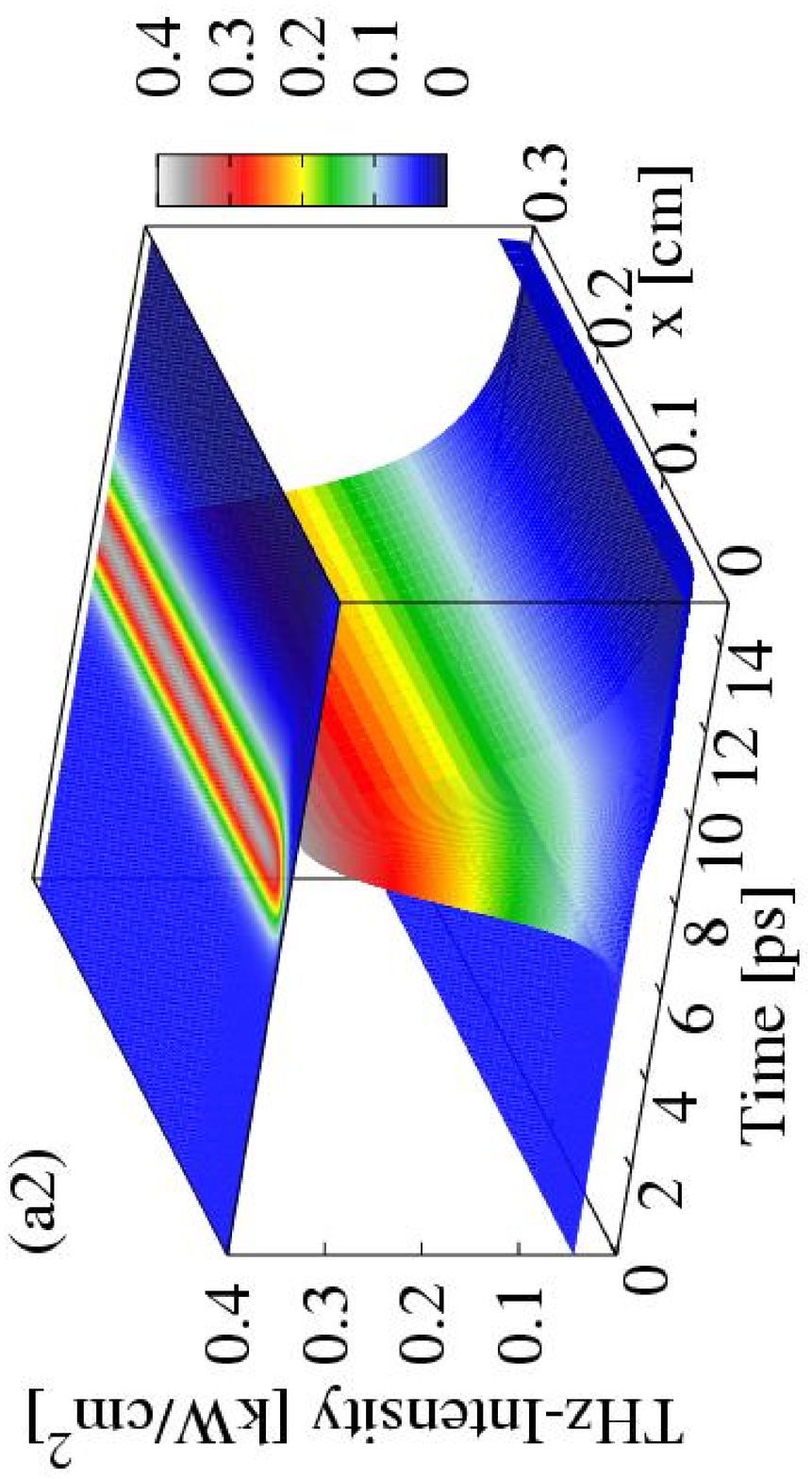}\hspace*{0,cm}
\includegraphics[width=0.43\columnwidth,angle=270]{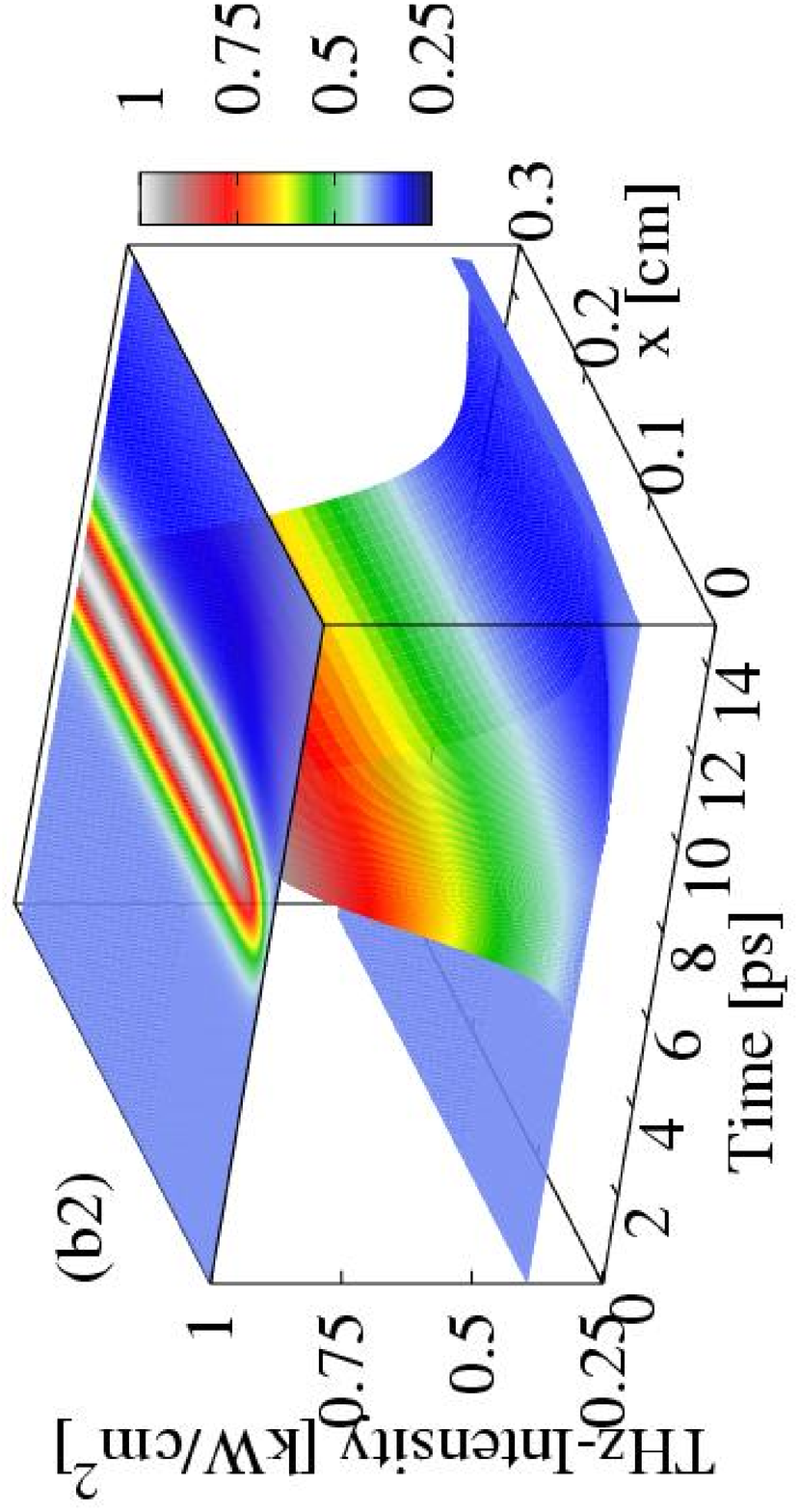}\hspace*{0.cm}
\includegraphics[width=0.43\columnwidth,angle=270]{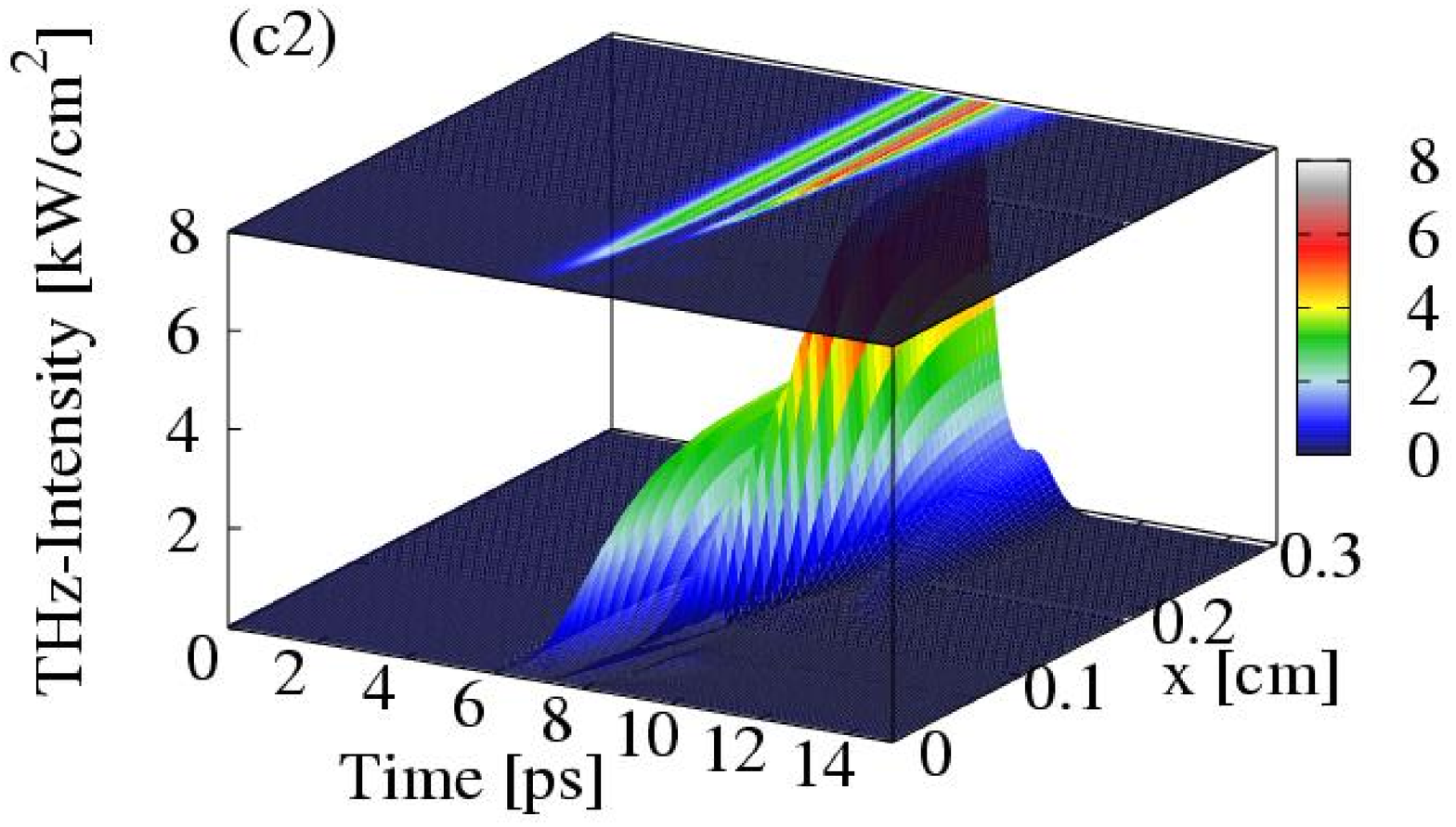}   
\caption{Color online: Spatial and temporal dependence of (1) population inversion, (2) generated THz-radiation for different combinations of drive and THz intensities illustrating the different origins of THz gain: stimulated emission (a) and quantum coherence effects (b,c).\vspace*{-0.8cm}\\ }
\end{figure*}

In this letter, we investigate the proposed optically-assisted electronically-driven scheme for a structure with a carrier density of $n = n_1 = \sum_i n_i = 5 \times 10^{10}$ cm$^2$. At lattice temperature T$_{\text{L}}=150$ K, carrier-carrier and carrier-phonon scattering redistribute the carriers between the subbands with $\gamma_{12}=0.5$ ps $^{-1}$, $\gamma_{23}=1.6$ ps$^{-1}$ and $\gamma_{34}=0.14$ ps$^{-1}$. These rates are determined from quantum kinetic calculations as described in Ref. \cite{we:IEEE}]. The final subband is depleted with an extraction rate of $\gamma_{\text{exc}} = 0.5 $ps$^{-1}$. 
In order to determine the gain of THz radiation, we probe the structure with a continuous wave (cw) THz-field.   
By optically exciting  the structure with a Gaussian pulse (temporal width $\sigma = 1$ ps, centered around $7.5$ ps) which spatially propagates in the quantum-well in-plane direction $x$, we control the population of the energetically higher laser subband 2. Intersubband transitions occurring between subband 2 and 3 yield the desired THz gain. 

Before we evaluate the benefit of recycling the pump photons in terms, we analyze the underlying physical effects for three 
different combinations of external excitation strength and probe intensity: (a) I$_{\text{max, D}}=2.8$ kW/cm$^{-2}$, I$_{\text{THz}}=0.044$ kW/cm$^{-2}$; (b) I$_{\text{max, D}}=2.8$ kW/cm$^{-2}$, I$_{\text{THz}}=0.394$ kW/cm$^{-2}$; (c) I$_{\text{max, D}}=26.5$ kW/cm$^{-2}$, I$_{\text{THz}}=0.044$ kW/cm$^{-2}$. As we will show in the following, these three cases highlight the importance of the relative strength of the radiative transitions and their connection to the two different origins of THz gain, i.e. stimulated emission and quantum coherence effects. 

In case (a), both the drive and the probe fields are too weak to stimulate substantial quantum coherence effects. The dipole-allowed polarizations are dominated by directly driven intersubband transitions (stimulated emission or absorption). During the first half of the drive pulse, the excitation of carriers from subband 1 into 2 and depletion of subband 3 via stimulated emission by the external optical field occurs on a faster timescale than the actual laser transition yielding strong population inversion [see Fig. 2(a1)], which drives the generation of THz radiation [see Fig. 2(a2)]. With decreasing drive pulse intensity, the strength of the radiative transitions between subband 1 and 2, and 3 and 4, is diminished. Radiative and non-radiative transitions between the laser subbands now occur faster than the radiative transitions between subbands 1 and 2. Furthermore, the decrease of drive intensity also yields an attenuation of the directly driven 3$\rightarrow$4 transitions, which are important for emptying the lower laser subband and regaining the pump intensity. Consequently, the lasing population inversion disappears.

In case (b), we use the same driving field as in case (a) but increase the probe intensity. Therewith, we increase the strength of the radiative transitions between the laser subbands. Intersubband transitions between the laser subbands occur faster than intersubband transitions between subband 1 and 2, and 3 and 4, respectively, population inversion does not occur [see Fig. 2(b1)]. Consequently, THz radiation cannot be generated via stimulated emission but only via quantum coherence effects. Whereas in case (a) quantum coherence effects were almost negligible, the increased probe intensity in case (b) allows the build-up of substantial quantum coherence contributions - substantial enough to create THz radiation [see Fig. 2 (b2)]. Cases (a) and (b) demonstrate clearly the importance of the relative strength of the radiative transitions. By changing their ratio externally (via an increase of the probe intensity) we were able to completely switch between the two origins of THz gain. 

However, in a QCL the THz generation will build up gradually. As a result, the relative strength of the radiative transitions and the magnitude of quantum coherence contributions will also change with time. THz radiation will first be generated only via stimulated emission, with increasing THz intensity quantum coherence effects will become more important. The latter can be seen in case (c), where we used the same low probe intensity as in case (a) but increased the driving field drastically. As in case (a), the relative strength of the radiative transitions allows the build-up of population inversion between the laser subbands. The THz field is too weak to create substantial quantum coherence contributions, stimulated emission is the cause of the observed gain in THz radiation. However, compared to case (a) we now excite the structure with a much stronger driving pulse, i.e. the driving pulse does not get absorbed as quickly as in case (a) and can propagate further into the structure. With increasing THz intensity (due to stimulated emission) and decreasing driving field, the relative strength of the radiative transitions changes. After the driving pulse has propagated roughly $0.15$ cm, the population inversion has almost completely vanished [see Fig. 2(c1)]. Stimulated emission is no longer the cause of the observed gain in THz radiation [see Fig. 2(c2)]. Due to the high intensity of generated THz radiation, quantum coherence effects are substantial enough to yield a further increase in THz radiation. 
Thus, whereas cases (a) and (b) demonstrate how an external increase in the probe field [from case (a) to (b)] can yield a switch between the two origins of THz gain, example (c) shows how for the case of a strong external drive field the same switch occurs gradually in the sample.
\begin{figure*}
\vspace*{-2cm}$\:$\\\includegraphics[width=0.8\columnwidth,angle=270]{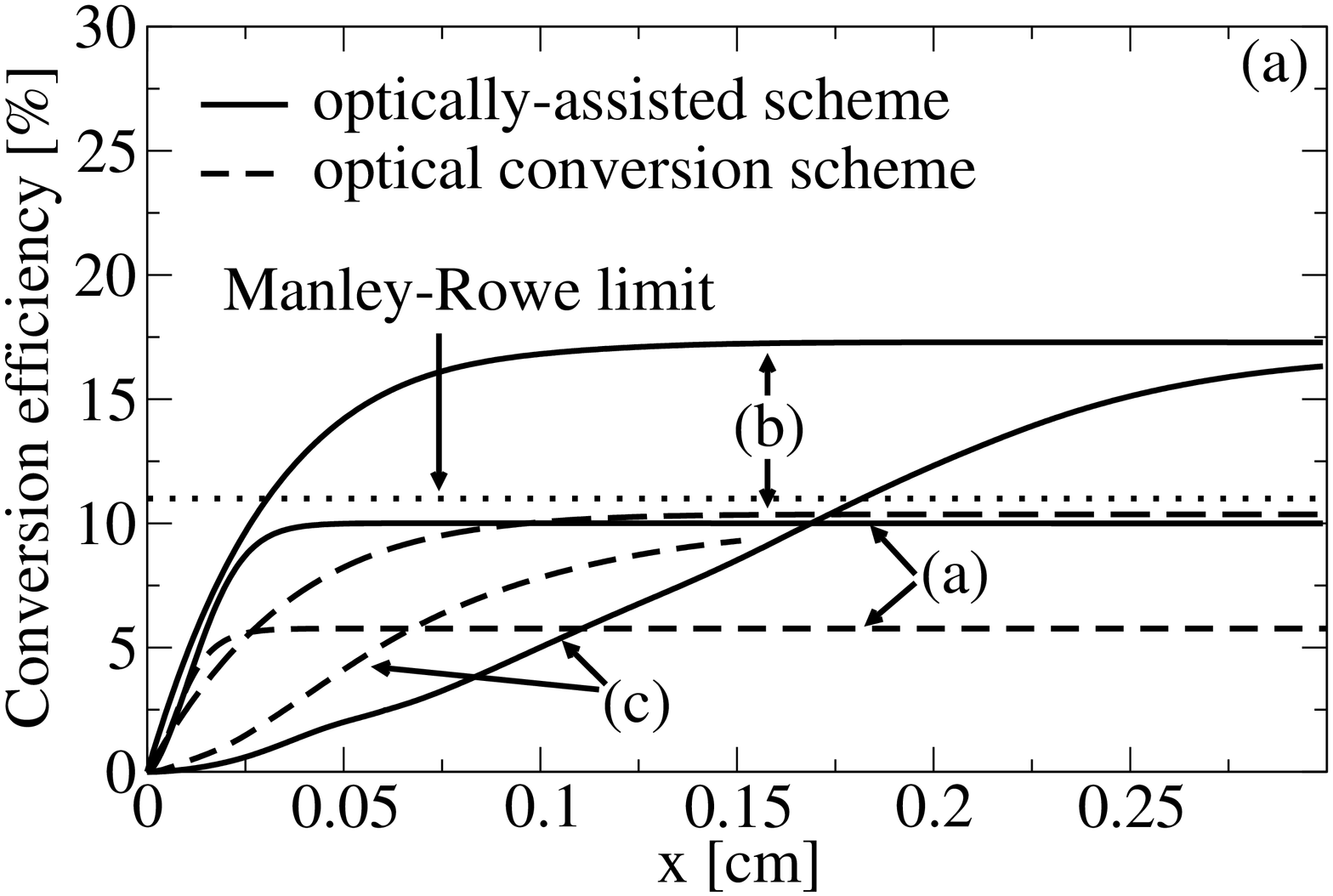}\includegraphics[width=0.8\columnwidth,angle=270]{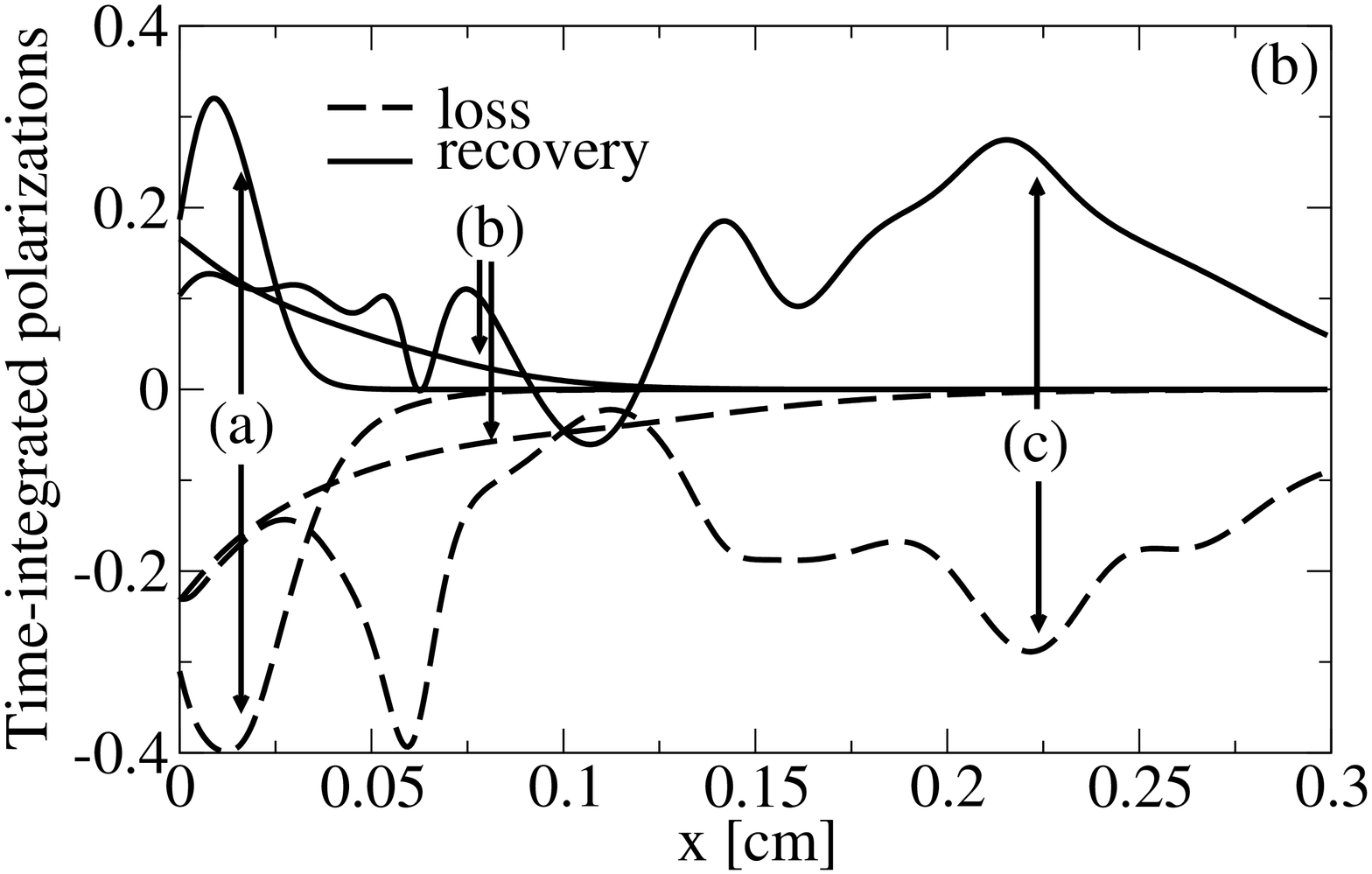}
\caption{(a) Comparison of the conversion efficiency between the optically-assisted scheme (solid lines) and the conventional optical conversion scheme (dashed lines) for the discussed combinations of drive and probe intensities (a,b,c). Whereas optical conversion schemes are limited by the Manley-Rowe limit (dotted line), optimization of the optical transition rates can yield optical efficiencies exceeding Manley-Rowe in the optically-assisted scheme. (b) Time-integrated polarizations as a measure for the success of recovery for cases (a,b,c) showing the changes in pump energy due to absorption and stimulated emission.\vspace*{-0.8cm}\\ }
\end{figure*}

To evaluate the benefit of recycling the pump photons, we present in Fig. 3(a) the conversion efficiencies for cases (a)-(c) and compare the results for the optically-assisted scheme and a conventional optical conversion scheme (no recovery). As can be seen, the optically-assisted scheme yields a substantial increase in final conversion efficiency due to the recovery of the drive field. The amount of recovery is shown in Fig. 3(b) in form of the imaginary part of the time-integrated polarizations, $\text{Im}(P_{21}(x))= - \int_0^{t_e} dt \: \text{Im}(P_{12}(x,t)) $ and $\text{Im}(P_{43}(x))= - \int_0^{t_e} dt \: \text{Im}(P_{34}(x,t)) $ with $t_e=15$ ps.. $\text{Im}(P_{21}(x))$ accounts for the relative loss of pump energy due to the excitation of carriers from the injection subband into the higher laser subband, $\text{Im}(P_{43}(x))$ gives the relative recovery of pump energy due to stimulated emission between the lower laser subband and subband 4. Comparing the relative strength of the time-integrated polarizations for the three cases shows that the recovery of the pump is more successful in case (b) than in case (a) which accounts for stronger improvement in the final conversion efficiency compared to the optical conversion scheme. In cases (b) and (c), the recovery of the pump energy is strong enough to yield conversion efficiencies clearly exceeding the Manley-Rowe relation. A complete recovery of the driving field intensity would yield a conversion efficiency of infinity. However, non-radiative transitions and damping of the polarizations due to carrier-carrier and carrier-phonon scattering reduce the recovery of the drive intensity.
Note that in case (c), the conventional optical conversion scheme at early distances actually yields a better efficiency than the optically-assisted scheme. This is because at early distances, the lower laser subband is depleted less efficiently in the optically-assisted scheme than in the optical conversion scheme yielding less THz-radiation via stimulated emission.    

Figure 3(a) also shows clearly the dependence of the conversion efficiency on the dominant origin of THz gain. Switching from stimulated emission to quantum coherence effects as the dominant contribution yields a dramatic increase in the achievable conversion efficiency [compare case(a) to case (b)] - for both the optically assisted approach and the optical conversion scheme. As the stimulated emission of THz radiation depends linearly on the population inversion, the conversion efficiency in case (a) is strongly suppressed by the amount of non-radiative transitions between the lasing subbands. The magnitude of the quantum coherence contributions on the other hand is less sensitive to the lifetimes of carriers in the higher laser subband yielding higher conversion efficiencies.   

In summary, we investigated the feasibility of an optically-assisted electrically driven THz scheme for THz-QCL. 
The uniqueness of the approach  involves the possibility of reducing the amount of parasitic scattering by energetically lifting the laser subbands, recycling the pump photons back into the system, and controlling conversion efficiency by tailoring the optical pump. For the example of pulse excitation, we showed the intricate interplay of stimulated emission and quantum coherence effects leading to efficient generation of THz radiation exceeding the Manley-Rowe quantum limit.

This work is funded by the US Department of Energy under contract DE-AC04-94AL8500 and the Alexander von Humboldt Foundation. 








\begin{thebibliography}{99}
 \bibitem{kohler:2002}
R. Koehler, et.al., Nature, \textbf{417}, 156 (2002)

\bibitem{williams:2003}
B. S. Williams, et al.,  Appl. Phys. Lett., \textbf{82}, 1015 (2003)
\bibitem{williams:2006}
B. S. Williams, et al.,  Appl. Phys. Lett., \textbf{88}, 261101 (2006)

\bibitem{kumar:2004}
 S. Kumar, et al., Appl. Phys. Lett., \textbf{84}, 2494 (2004);

\bibitem{gauthier:1999}
O. Gauthier-Lafaye, et al., Appl. Phys. Lett., \textbf{74}, 1537 (1999)

\bibitem{liu_raman:2003}
 H.C. Liu, et.al., Phys. Rev. Lett., \textbf{90}, 077402 (2003)

\bibitem{we:IEEE}
I. Waldmueller et al., IEEE JQE, \textbf{42}, 292 (2006)
\end{thebibliography}
\end{document}